\documentclass{ifacconf}

\usepackage{graphicx}      
\usepackage{natbib}        
\usepackage{amsmath}
\usepackage{amssymb}
\usepackage{xcolor}

\begin{document}
\begin{frontmatter}

\title{Multi-Domain Graph-Based Modeling of Energy Systems with Applications to Lithium-Ion Batteries} 


\author[First]{Mahsa Hemmat} 
\author[First]{Andrew G. Alleyne} 

\address[First]{Department of Mechanical Engineering, University of Minnesota, Minneapolis, MN 55455 USA (e-mail: hemma031@umn.edu, alleyne@umn.edu).}

\begin{abstract}                
Graph-based models have been shown to provide a structured representation for complex multi-domain energy systems but face limitations when edge power flows depend on non-adjacent states or when a single edge carries multiple power-flow types driven by different inputs. This paper proposes two general extensions to address these limitations: a recursive state-to-input feedback scheme that embeds non-adjacent state dependencies into edge inputs without altering the graph structure, and a parallel edge decomposition method that represents composite interactions using sets of single-input edges while preserving energy conservation at the vertices. The extended framework is demonstrated on a lithium-ion battery module consisting of 36 parallel cells, and the resulting model predicts module temperatures with errors below 1°C. Validation on this electro-thermal battery system demonstrates the effectiveness of the extended framework for multi-domain systems that cannot be represented by previously established graph-based formulations, and indicates its potential for broader application to complex energy systems in control and design studies.
\end{abstract}

\end{frontmatter}

\begingroup
\renewcommand\thefootnote{}\footnotetext{\copyright~2026 the authors. This work has been accepted to IFAC for publication under a Creative Commons Licence CC-BY-NC-ND.}\addtocounter{footnote}{0}
\endgroup

\section{Introduction}


{Modern energy systems, such as electrified vehicles, are large interconnected systems composed of distinct components --- battery packs, motors, power electronics, and thermal management subsystems --- that couple electrical, thermal, and mechanical domains and therefore involve strong cross-domain interactions} \citep{Williams:18,Koeln:20}. {Capturing the behavior of such systems with sufficient fidelity for estimation, monitoring, and control requires models that span multiple physical domains simultaneously and scale to the large number of states that result from their interconnected structure} \citep{Aksland:19}. {While conventional state-space models can represent such dynamics, the complexity, multi-domain nature, and modular physical construction of these systems motivate a modeling framework that goes beyond a single monolithic equation set.}

{Graph-based models provide such a modeling framework by representing system components as vertices and their physical interactions as edges, with each edge encoding a power exchange between its connected vertices formulated to satisfy conservation of energy~\citep{Koeln:16}. While the graph-based model ultimately takes the form of an ordinary differential equation (ODE) system, it is more than a representational choice, as the graph structure itself offers additional capabilities that a state-space formulation does not naturally provide. The framework is inherently modular, where each subsystem of a large energy system, for example an electric motor or a converter in an electrified vehicle, can be modeled and validated separately and then used as a reusable graph-based component. A full system model is then built by interconnecting these components through edges that capture their couplings across electrical, thermal, and mechanical domains; this process is referred to as scalable composition~\citep{Park:23}. This scalable composition and its systematic nature reduce implementation errors that commonly occur in manual large-scale formulations, such as inconsistent sign conventions, missing coupling terms, or index mismatches. The component-based structure of the framework further supports a plug-and-play mode of operation in which alternative system architectures and component configurations can be evaluated by modifying only the inter-component couplings, leaving the internal dynamics of each component unchanged, which makes it well-suited for systematic design and control evaluation. In addition to these benefits, the graph structure also enables the use of graph theory techniques for formal analysis of system interconnections, supporting applications such as model decomposition and hierarchical control design} \citep{Pisani:26}.

These features have enabled graph-based modeling to be applied across a broad range of multi-domain energy systems, including aircraft electro-thermal systems \citep{Williams:18,Koeln:20,Aksland:22}, liquid-cooling and thermal-fluid networks \citep{Yang:18,Lionello:20,Russell:22}, high-pressure hydraulic networks \citep{Niederberger:18}, and electro-thermal components in electric vehicles \citep{Docimo:18}. {To support these applications, an open-source MATLAB/Simulink toolbox has been developed to facilitate the generation and analysis of graph-based models across multiple energy domains, and is publicly available at the GitHub repository \texttt{psu-PAC-Lab/graph-model-toolbox}} \citep{Pisani:26}.

{Despite this broad applicability, current graph-based formulations have limitations that restrict their use in a wider class of multi-domain energy systems.} In existing graph-based formulations, an edge power flow is written as a function of variables at its head and tail vertices and explicit edge inputs \citep{Koeln:16,Aksland:21}. The first limitation occurs when an edge depends on a state in a vertex that is neither its head nor its tail, referred to here as a non-adjacent vertex. In that case, the framework lacks a standard mechanism to express this cross-vertex dependence within the existing graph-based formulation.

Also, in current graph-based literature, an edge typically has a single input~\citep{Koeln:16,Aksland:21}. In practice, however, an edge may deliver the sum of distinct power-flow types, each governed by a different input. Therefore, a method is needed that enables such edges while preserving the original dynamics and exact energy balance at the vertices and maintaining consistency with the graph formulation. These two limitations are common in coupled multi-domain energy systems (for example, electro-thermal systems), yet prior graph-based studies have not provided a general, systematic solution.

This paper addresses these limitations by introducing two general modeling strategies for graph-based energy-system models. First, a recursive state-to-input feedback scheme is formulated that allows edge power flows to depend on states in non-adjacent vertices while remaining consistent with the prior graph-based formulation. Second, a parallel edge decomposition is proposed for composite edges, in which distinct power-flow types with different inputs are represented by separate edges whose contributions sum at the vertex, preserving the original dynamics and exact energy balance.

The remainder of this paper is organized as follows.
Section~\ref{sec:graph_modeling} introduces the fundamental graph-based modeling framework and then extends it with two methods, the recursive state-to-input feedback method and the parallel edge decomposition method, that address the limitations discussed earlier. Section~\ref{sec:validation_module} applies the proposed methods to a Li-ion battery module, presents its graph-based formulation, and compares model predictions with experimental temperature measurements.
Section~\ref{sec:conclusion} concludes the paper by summarizing the main contributions and outlining directions for future research.

\section{Graph-based Modeling Framework}
\label{sec:graph_modeling}
\subsection{Graph-based Modeling Fundamentals}
\label{sec:graph_fundamentals}

\begin{figure}[t]
    \begin{center}
        \includegraphics[width=\columnwidth]{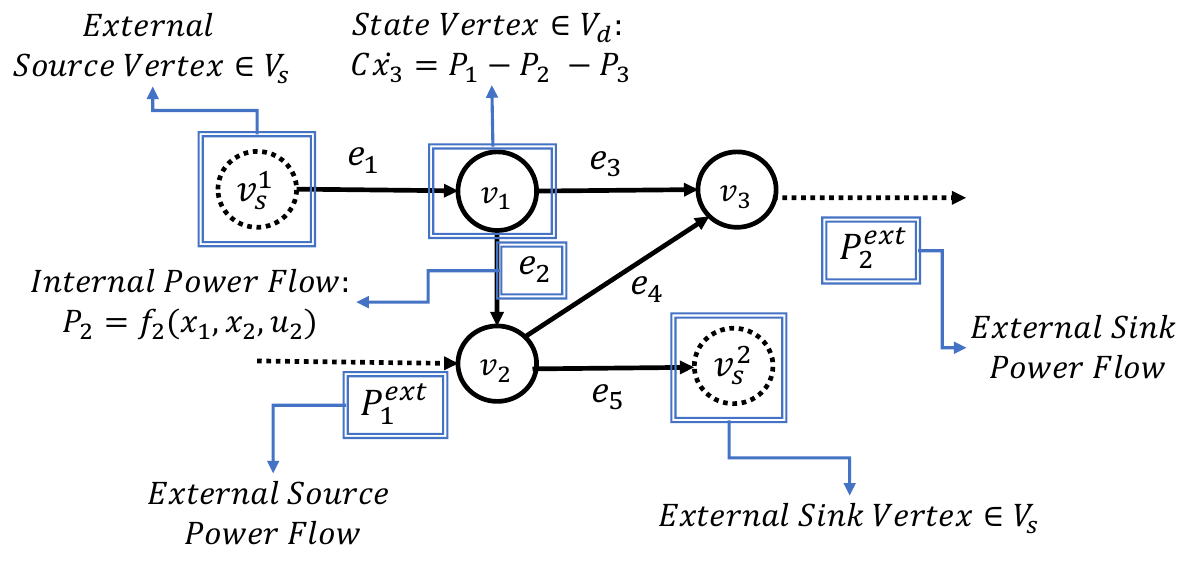}
        \caption{Generic example of a graph-based model}
        \label{fig:graph_overview}
    \end{center}
\end{figure}

Graphs provide a general mathematical framework for representing various types of interconnected systems.
In this work, an oriented graph $G = (V, E)$ with vertex set $V = \{v_i\}_{i=1}^{N_v}$ and edge set
$E = \{e_j\}_{j=1}^{N_e}$ is used to represent an energy system, where vertices correspond
to energy storage elements and edges represent power exchange interactions, called power flows, between them.
Each edge is oriented from its tail vertex to its head vertex, defined as positive power flow by convention.
Figure~\ref{fig:graph_overview} shows a generic example of such a graph model.

{The vertex set $V$ is partitioned into two subsets: dynamic vertices $V_d$ and external
sink/source vertices $V_s$, so that $V = V_d \cup V_s$ with $N_d = |V_d|$ and $N_s = |V_s|$.
Each dynamic vertex $v_i \in V_d$ stores energy and, in the single-state vertex formulation, carries a scalar
state $x_i$ with an associated capacitance $C_i > 0$; these states are modeled and solved for within the framework, and are collected in the full state vector $x \in \mathbb{R}^{N_d}$.
The physical meaning of $C_i$ and $x_i$ depends on the underlying subsystem's domain: electrical}
vertices use capacitors or inductors with voltage or current states, mechanical vertices use masses or inertias
with velocity or angular speed states, and thermal vertices use thermal capacitances with temperature
states; further details on this mapping can be found in \citet{Aksland:21}.
This paper adopts the single-state {vertex} formulation{, in which each
dynamic vertex is associated with a single scalar state $x_i$}; an extended graph-based formulation that
allows each vertex to store a vector of states, i.e.\ multi-state vertices, is described
in~\citet{Russell:22}, but it is not considered further here.

{External sink/source vertices in $V_s$, each associated with a scalar variable $x_i^s$,
do not store energy. They serve as boundary interfaces through which the system exchanges power with its
surroundings, and their quantities are specified externally, typically by neighboring subsystems or the
ambient environment, and treated as disturbances to the system rather than solved for within the model.}

{The edges in $E$ are grouped into two categories: internal edges $E_{\mathrm{int}}$ and
external edges $E_{\mathrm{ext}}$, so that $E = E_{\mathrm{int}} \cup E_{\mathrm{ext}}$ with
$N_{\mathrm{int}} = |E_{\mathrm{int}}|$ and $N_{\mathrm{ext}} = |E_{\mathrm{ext}}|$.
Each internal edge $e_j \in E_{\mathrm{int}}$ carries a power flow $P_j$ within the system, determined by}
the states of its adjacent vertices and a corresponding edge input.
For an internal edge $e_j$, the power flow is written in general as
\begin{equation}
    P_j = f_j\!\left(x_j^{\mathrm{tail}},\, x_j^{\mathrm{head}},\, u_j\right),
    \label{eq:generic_edge}
\end{equation}
where $f_j(\cdot)$ is an arbitrary nonlinear function of the tail and head states and a scalar input $u_j$.
{The internal power flows are collected in the vector $P \in \mathbb{R}^{N_{\mathrm{int}}}$.}

{External edges represent external power flows and are collected in a vector
$P^{\mathrm{ext}} \in \mathbb{R}^{N_{\mathrm{ext}}}$.} These flows represent exchanges with the surroundings
and are specified directly by the environment or neighboring subsystems. They are not functions of the system
states and can be computed outside the model and supplied to the graph-based framework as disturbances.
{The different edge types and their connections to internal and external source/sink vertices
are illustrated schematically in Fig.~\ref{fig:graph_overview}.}

{With the vertex sets and edge categories defined, applying} conservation of energy at a
{dynamic} vertex $v_i \in V_d$ {then} yields the local state dynamics:
\begin{equation}
    C_i \dot{x}_i
    =
    \sum_{e_j \in E_{i,\mathrm{int}}^{\mathrm{head}}}\!\!P_j
    -
    \sum_{e_j \in E_{i,\mathrm{int}}^{\mathrm{tail}}}\!\!P_j
    +
    \sum_{e_k \in E_{i,\mathrm{ext}}^{\mathrm{head}}}\!\!P_k^{\mathrm{ext}}
    -
    \sum_{e_k \in E_{i,\mathrm{ext}}^{\mathrm{tail}}}\!\!P_k^{\mathrm{ext}},
    \label{eq:vertex_dynamics}
\end{equation}
where $E_{i,\mathrm{int}}^{\mathrm{head}}$ and
$E_{i,\mathrm{int}}^{\mathrm{tail}}$ denote the sets of internal edges
entering and leaving the vertex, and $E_{i,\mathrm{ext}}^{\mathrm{head}}$ and
$E_{i,\mathrm{ext}}^{\mathrm{tail}}$ represent the corresponding sets for external edges.

The connectivity between vertices and internal edges is encoded in an incidence matrix
$M = [m_{ij}]$, formed over the vertices in $V_d \cup V_s$ and the internal edges
in $E_{\mathrm{int}}$, with dimension $(N_d + N_s) \times N_{\mathrm{int}}$.
Each entry $m_{ij}$ indicates how internal edge $e_j$ is oriented with respect to vertex $v_i$:
$m_{ij} = 1$ if $v_i$ is the tail of $e_j$, $m_{ij} = -1$ if $v_i$ is the head of $e_j$, and $m_{ij} = 0$
otherwise. With the dynamic vertices in $V_d$ ordered first and the
sink/source vertices in $V_s$ following, the incidence matrix is written in block form
\begin{equation}
    M =
    \begin{bmatrix}
        \overline{M} \\
        \underline{M}
    \end{bmatrix},
\end{equation}
where $\overline{M} \in \mathbb{R}^{N_d \times N_{\mathrm{int}}}$ contains the rows
associated with dynamic vertices and maps internal edge flows to the state dynamics, and
$\underline{M} \in \mathbb{R}^{N_s \times N_{\mathrm{int}}}$ contains
the rows associated with the sink/source vertices.

External power flows are mapped to the dynamic vertices
through an interconnection matrix $D = [d_{ij}] \in \mathbb{R}^{N_d \times N_{\mathrm{ext}}}$.
The entry $d_{ij}$ describes how external flow $P^{\mathrm{ext}}_j$ is connected to a
dynamic vertex $v_i \in V_d$: $d_{ij} = 1$ if $v_i$ is the head of the external flow,
$d_{ij} = -1$ if $v_i$ is the tail, and $d_{ij} = 0$ otherwise.

Using these interconnection matrices, the dynamics of the graph-based model can be written in compact form as
\begin{equation}
    C \dot{x} = -\,\overline{M}\,P(x, x^s, u) + D P^{\mathrm{ext}},
    \label{eq:compact_dynamics}
\end{equation}
where $C = \mathrm{diag}(C_i)$ is the diagonal matrix of capacitances, $x \in \mathbb{R}^{N_d}$ is the
state vector, $x^s \in \mathbb{R}^{N_s}$ collects the variables associated
with the sink/source vertices, $u$ is the vector of edge inputs,
$P(x, x^s, u) \in \mathbb{R}^{N_{\mathrm{int}}}$ is the vector of
internal power flows defined by the edge law in~\eqref{eq:generic_edge}, and
$P^{\mathrm{ext}} \in \mathbb{R}^{N_{\mathrm{ext}}}$ is the vector of
external power flows.

\subsection{Recursive State-to-Input Feedback Method}
In the existing graph formulation, each internal power flow is written as a function of the adjacent vertex states and an edge input, as in~\eqref{eq:generic_edge}, but in multi-domain models some power flows also depend on non-adjacent states. A typical example is an electrical or mechanical edge whose parameters vary with a thermal state elsewhere in the graph. In these cases, the physical law is written in the form \(P_j = f_j\bigl(x_j^{\text{tail}}, x_j^{\text{head}}, g(x_\ell)\bigr)\), where \(x_\ell\) is a non-adjacent state and \(g(\cdot)\) represents how that state influences the edge. The mapping \(g(\cdot)\) may return the state itself (e.g., \(g(x_\ell) = x_\ell\)) or a property derived from it, such as a temperature-dependent resistance. Such non-adjacent dependencies do not fit the existing edge structure, which is restricted to adjacent states and edge inputs, and therefore cannot be represented directly within the usual formulation.

The recursive state-to-input feedback method restores the previously established edge structure while retaining these non-adjacent dependencies. A selected non-adjacent state is exposed as an output of the graph model, \(y = x_\ell\), mapped through \(g(\cdot)\) to define the edge input \(u_j = g(y) = g(x_\ell)\), and then used in the existing edge law \(P_j = f_j\bigl(x_j^{\mathrm{tail}}, x_j^{\mathrm{head}}, u_j\bigr)\). In this way, the dependence on the non-adjacent state \(x_\ell\) is carried through the edge input rather than as an additional state argument.

In this method, the feedback is purely algebraic and internal to the model's signal structure, so it does not introduce any new dynamics or control law and the state dynamics remain unchanged. 

\begin{figure}[!t]
    \centering
    \includegraphics[width=0.9\columnwidth]{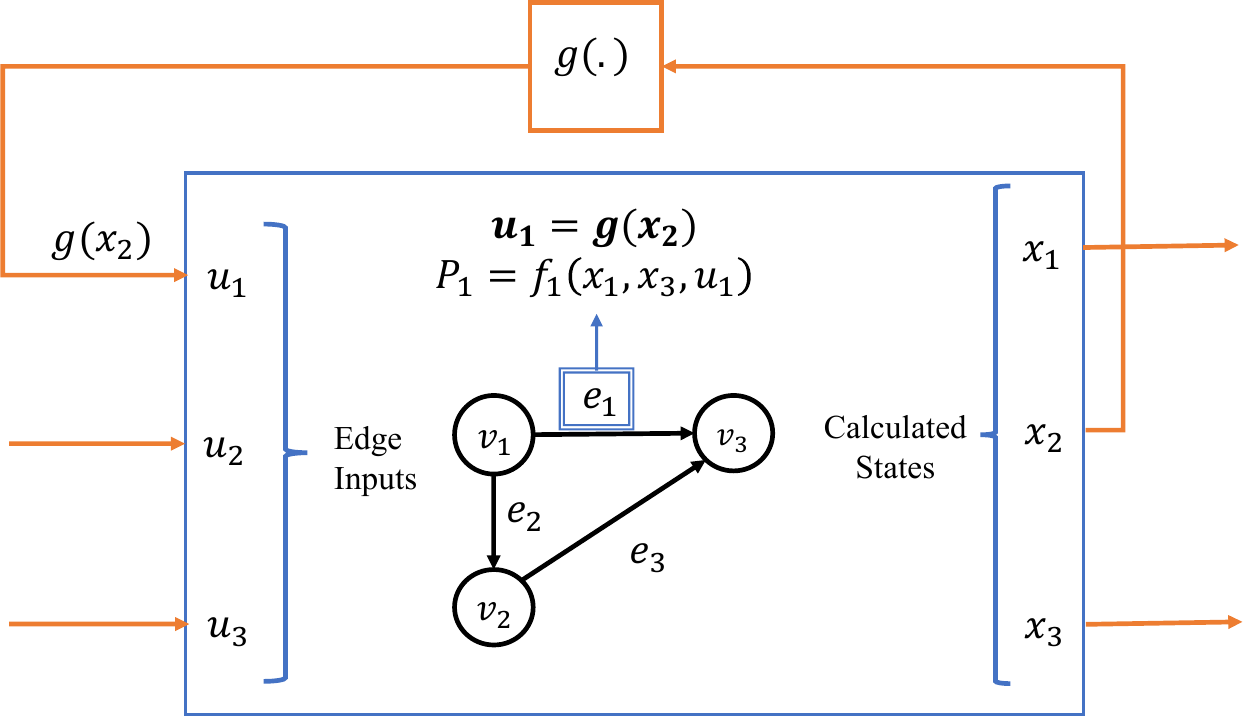}%
    \caption{Recursive state-to-input feedback method within a graph-based model.}
    \label{fig:feedback_method}
\end{figure}

Fig.~\ref{fig:feedback_method} illustrates this construction. The blue block represents the internal graph model with state vector $x$ and internal power flows $P$. An internal edge between vertices $v_1$ and $v_3$ is influenced by a non-adjacent vertex $v_2$: its state is read out as $y = x_2$, mapped through $g(\cdot)$ to form the edge input $u_j = g(x_2)$, and then used in the edge law. The edge therefore still appears only between $v_1$ and $v_3$ in the graph and incidence matrix, while its power flow depends on the remote state $x_2$ through the state-to-input feedback loop.

\subsection{Parallel Edge Decomposition Method}

In the existing graph formulation, although the internal edge law~\eqref{eq:generic_edge} with function $f_j(\cdot)$ can represent any nonlinear behavior, it is often convenient for analysis and implementation to express the edge law using the aggregated polynomial form introduced in~\citet{Aksland:21}. In this representation, each power flow $P_j$ is written as
\begin{equation}
\begin{aligned}
    P_j
    &=
    g_j(x_t, x_h, u_j)
    \bigl(
        c_1 x_t
      + c_2 x_h
      + c_3 x_h x_t
      + c_4 x_t^2        \\
    &\qquad\qquad
      + c_5 x_h x_t^2
      + c_6 x_h x_t u_j
      + c_7 x_t^2 u_j
      + c_8 x_t^3
    \bigr)
    \label{eq:aggregated_edge}
\end{aligned}
\end{equation}
where $x_t = x_j^{\mathrm{tail}}$ and $x_h = x_j^{\mathrm{head}}$ are the tail and head vertex states, $c_k$ are constant aggregated coefficients, and $g_j(\cdot)$ is any scalar function of the adjacent vertex states and the edge input. In practice, $g_j$ is typically implemented as a lookup table to capture property variations that are difficult to represent analytically. {Every common power-flow interaction in thermal, electrical, and mechanical energy systems can be described by a single monomial or a combination of the monomials in~\eqref{eq:aggregated_edge}, and by selecting which coefficients $c_k$ are nonzero, this template reproduces a wide variety of common edge models across these domains; examples are reported in~\citet{Pisani:26}.}

A key observation from both \eqref{eq:generic_edge} and \eqref{eq:aggregated_edge} is that each edge is defined with exactly one input $u_j$. In some multi-domain systems, however, the physical power flow {along a single interaction} may naturally appear as a sum of {terms from the polynomial template~\eqref{eq:aggregated_edge},} {each driven by its own input.} For instance,
\begin{equation}
    e_{1}:~P_1 = {u_{1,1}}\,x_t + {u_{1,2}}\,x_t x_h,
\end{equation}
{is an abstract motivating example in which two power-flow types, $x_t$ and $x_t x_h$, are each driven by a separate input; a concrete physical realization of this structure appears in Section~\ref{sec:validation_module}.} The total flow cannot be {written as} $f_j(x_t, x_h, u_j)$ with {a single} effective input{, so} the interaction cannot be assigned to a single edge under{~\eqref{eq:aggregated_edge}.}

To represent such interactions while remaining consistent with the graph-based framework, this work introduces a parallel edge decomposition method. When the total power flow along a physical connection is a sum of {power-flow types from~\eqref{eq:aggregated_edge},} each driven by its own input, the interaction is modeled as a set of parallel edges {sharing} the same head and tail vertices. For the example above, the decomposition is
\begin{equation}
\begin{aligned}
    e_{1a}:~P_{1a} &= {u_{1,1}}\,x_t {= f_{1,1}(x_t, x_h, u_{1,1})}, \\
    e_{1b}:~P_{1b} &= {u_{1,2}}\,x_t x_h {= f_{1,2}(x_t, x_h, u_{1,2})}.
\end{aligned}
\label{eq:decomposition_example}
\end{equation}
The two edges connect the same pair of vertices and {produce} the same net effect on the vertex balances, but each {now conforms to the single-input edge structure~\eqref{eq:generic_edge}} and can be incorporated into the graph model without modifying the underlying formulation.

\begin{figure}[!t]
    \centering
    \includegraphics[width=0.95\columnwidth]{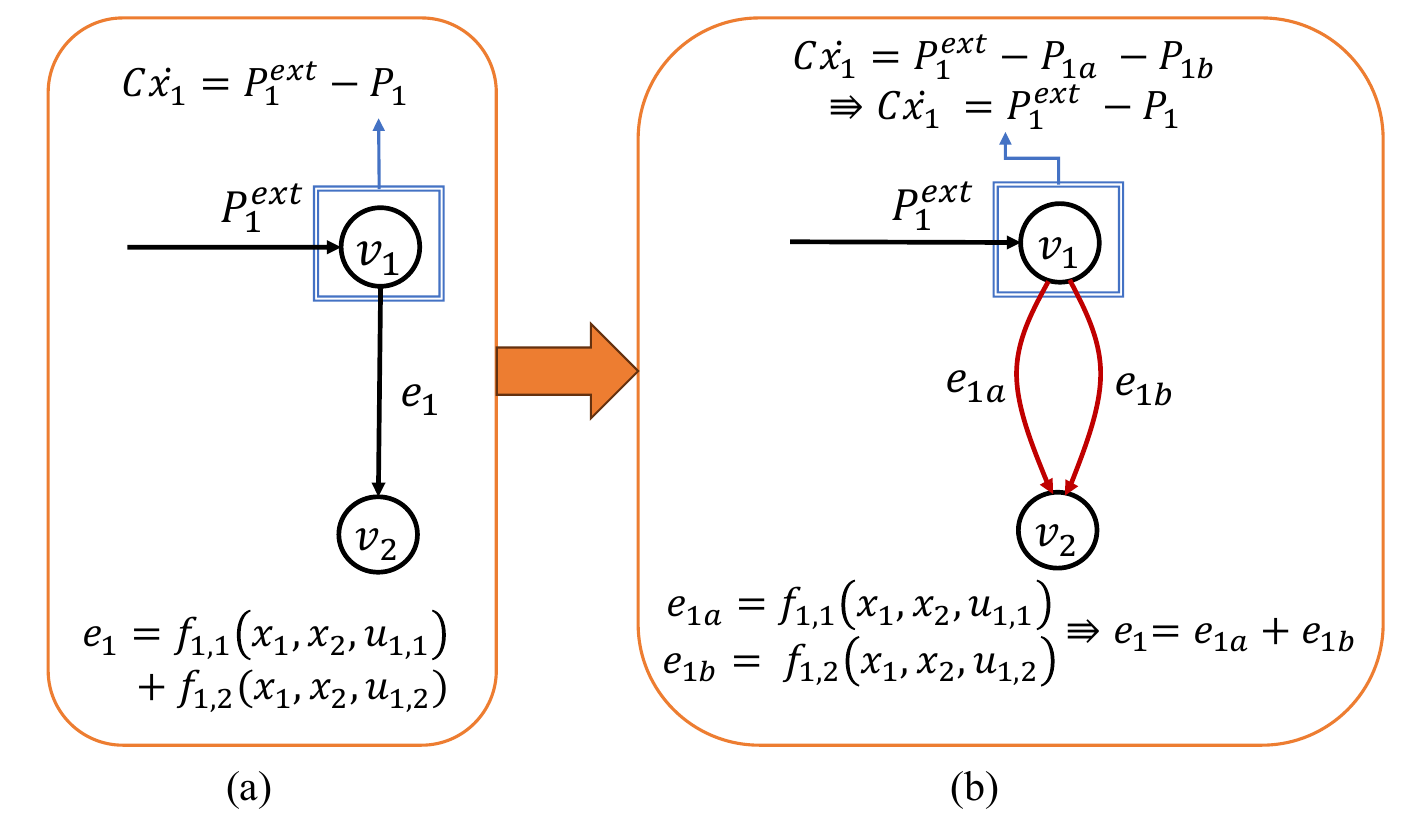}%
    \caption{Parallel-edge decomposition of a multi-input power flow within a graph-based model}
    \label{fig:parallel_edges}
\end{figure}

Fig.~\ref{fig:parallel_edges} illustrates the decomposition. The left panel shows a single power flow written as the sum of two structurally different terms acting through one physical interaction. The right panel shows the equivalent representation after replacing that interaction with two parallel edges sharing the same head and tail vertices, each carrying one power-flow type with a single assigned input following the edge law~\eqref{eq:generic_edge}. Because the total power into and out of each vertex is preserved, the vertex dynamics remain unchanged.

The parallel edge decomposition is particularly useful in combination with the recursive state-to-input feedback method. When that method introduces additional effective inputs for a single physical connection, the decomposition assigns each input to a separate parallel edge, separating the associated power-flow types while preserving the previously established graph-based structure.

\section{Validation of Graph-Based Modeling Methods on a Li-Ion Battery Module}
\label{sec:validation_module}
To evaluate the proposed modeling framework, a Li-ion battery system is selected and compared against experimental test data. This system is multi-domain and exhibits both non-adjacent state dependencies and composite heat-generation mechanisms, {precisely the limitations that the prior graph-based formulation cannot handle. Because the system cannot be represented within the prior formulation without the proposed extensions, it serves as a direct test case for evaluating the methods introduced in this work. Validation is performed by comparing model predictions against experimental measurements.}

\subsection{Module Architecture and Electro-Thermal Dynamics}
\label{subsec:module_dynamics}

\begin{figure}[!b]
    \centering
    \includegraphics[width=0.98\columnwidth]{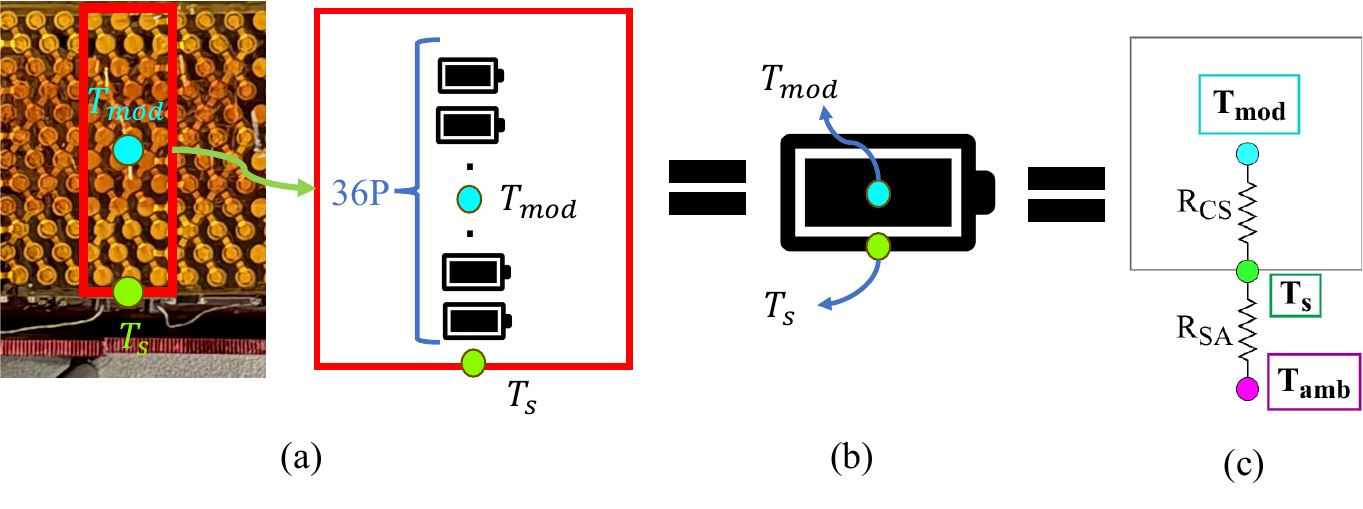}%
    \caption{SVP module layout and its equivalent thermal network}
    \label{fig:svp_module_photo}
\end{figure}

For experimental validation, the University of Minnesota Solar Vehicle Project (SVP) battery pack is used. This pack consists of blocks of 36 cells connected in parallel (36P), and 36 of these blocks connected in series (36S), resulting in an overall 36P36S architecture. Each 36P block is referred to as a module. For simplicity, the validation focuses on a single module. Fig.~\ref{fig:svp_module_photo}(a) shows the pack with one 36P module highlighted. Two temperature sensors were installed on this module: one placed near the geometric center of the 36P block to measure the internal module temperature $T_{\mathrm{mod}}$, and another mounted on the side board adjacent to the same block to measure a representative side temperature $T_s$. 

Following the framework in~\citet{Hemmat:26}, these two measurements motivate a two-node thermal network with one electrochemical node at $T_{\mathrm{mod}}$ and one structural node at $T_s$. Each 36P block is treated as a single lumped module with uniform behavior, as sketched in Fig.~\ref{fig:svp_module_photo}(b), and its thermal dynamics are captured by two first-order energy balances for the module and side-board nodes, using the equivalent thermal circuit shown in Fig.~\ref{fig:svp_module_photo}(c).
Because some properties in these thermal dynamics depend on state of charge (SOC), SOC is also modeled by the Coulomb counting method.
Thus, the governing equations are
\begin{equation}
    \dot{\mathrm{SOC}} = -\,\frac{I_{\mathrm{mod}}}{3600\,B_{\mathrm{cap,mod}}},
    \label{eq:soc_dynamics}
\end{equation}
\begin{equation}
\begin{split}
    C_{\mathrm{th,mod}}\dot{T}_{\mathrm{mod}}
    &=
    I_{\mathrm{mod}}\bigl( V_{\mathrm{ocv,mod}} - V_{T,\mathrm{mod}} \bigr)
    \\
    &\quad
    -\,I_{\mathrm{mod}}T_{\mathrm{mod}}
    \left(\frac{\partial V_{\mathrm{ocv}}}{\partial T}\right)_{\mathrm{mod}}
    +\,\frac{T_s - T_{\mathrm{mod}}}{R_{\mathrm{CS}}}
\end{split}
    \label{eq:Tmod_dynamics}
\end{equation}
\begin{equation}
    C_{\mathrm{th,s}} \dot{T}_s
    =
    \frac{T_{\mathrm{mod}} - T_s}{R_{\mathrm{CS}}}
    + \frac{T_{\mathrm{amb}} - T_s}{R_{\mathrm{SA}}}.
    \label{eq:Ts_dynamics}
\end{equation}
Here $B_{\mathrm{cap,mod}}$ is the usable capacity of the module (Ah), and $I_{\mathrm{mod}}$ is the measured module current (A, positive for discharge). The term $I_{\mathrm{mod}}\bigl(V_{\mathrm{ocv,mod}} - V_{T,\mathrm{mod}}\bigr)$ represents irreversible Joule heating, where $V_{\mathrm{ocv,mod}}$ is the open-circuit module voltage (V) as a function of SOC and $V_{T,\mathrm{mod}}$ is the measured terminal voltage (V). The term $I_{\mathrm{mod}} T_{\mathrm{mod}} \bigl(\partial V_{\mathrm{ocv}}/\partial T)_{mod}$ represents reversible entropic heating, where $(\partial V{_\mathrm{ocv}}/\partial T)_{mod}$ is the SOC-dependent entropic coefficient of the module (V/K). Both heating terms have units of watts. The parameters $C_{\mathrm{th,mod}}$ and $C_{\mathrm{th,s}}$ are the thermal capacitances (J/K) of the module and side nodes, respectively. The resistances $R_{\mathrm{CS}}$ and $R_{\mathrm{SA}}$ are the effective thermal resistances (K/W) between the module and side nodes and between the side node and ambient, respectively, accounting for conduction through the pack structure and natural convection with the surrounding air.


The structure of these equations matches the module-level formulation in earlier validated thermal modeling work on the same pack~\citep{Hemmat:26}, except that heat transfer between adjacent modules is neglected. This approximation is reasonable because the experiments were performed without active cooling and under nearly uniform ambient conditions, and all series modules are assumed identical so that they experience similar heating.

{Parameters are taken directly from~\citet{Hemmat:26}, as the present experiments use the same pack under similar loading and cooling conditions. For the 36P module ($N_{\mathrm{parallel}} = 36$, $N_{\mathrm{series}} = 1$), $B_{\mathrm{cap,mod}} = 162~\mathrm{Ah}$, $C_{\mathrm{th,mod}} = 2100~\mathrm{J/K}$, $C_{\mathrm{th,s}} = 35~\mathrm{J/K}$, $R_{\mathrm{CS}} = 40~\mathrm{K/W}$, and $R_{\mathrm{SA}} = 60~\mathrm{K/W}$. Because $N_{\mathrm{series}} = 1$, the module OCV and entropic coefficient are identical to the single-cell properties; their characterization follows~\citet{Hemmat:26}.}

\subsection{Graph-Based Formulation of the Battery Module}

To represent the battery module in the graph-based framework, all state balances must be written in power form. The temperature dynamics in \eqref{eq:Tmod_dynamics}--\eqref{eq:Ts_dynamics} already appear in this form. The SOC equation \eqref{eq:soc_dynamics} is rewritten in power form as
\begin{equation}
    {3600\,B_{\mathrm{cap,mod}}} V_{\mathrm{ocv,mod}}\,\dot{\mathrm{SOC}}
    =-{V_{\mathrm{ocv,mod}}\,I_{\mathrm{mod}}},
    \label{eq:soc_dynamics_power}
\end{equation}
which is used in place of \eqref{eq:soc_dynamics} for graph-based modeling.

\begin{figure}[!t]
    \centering
    \includegraphics[width=0.95\columnwidth]{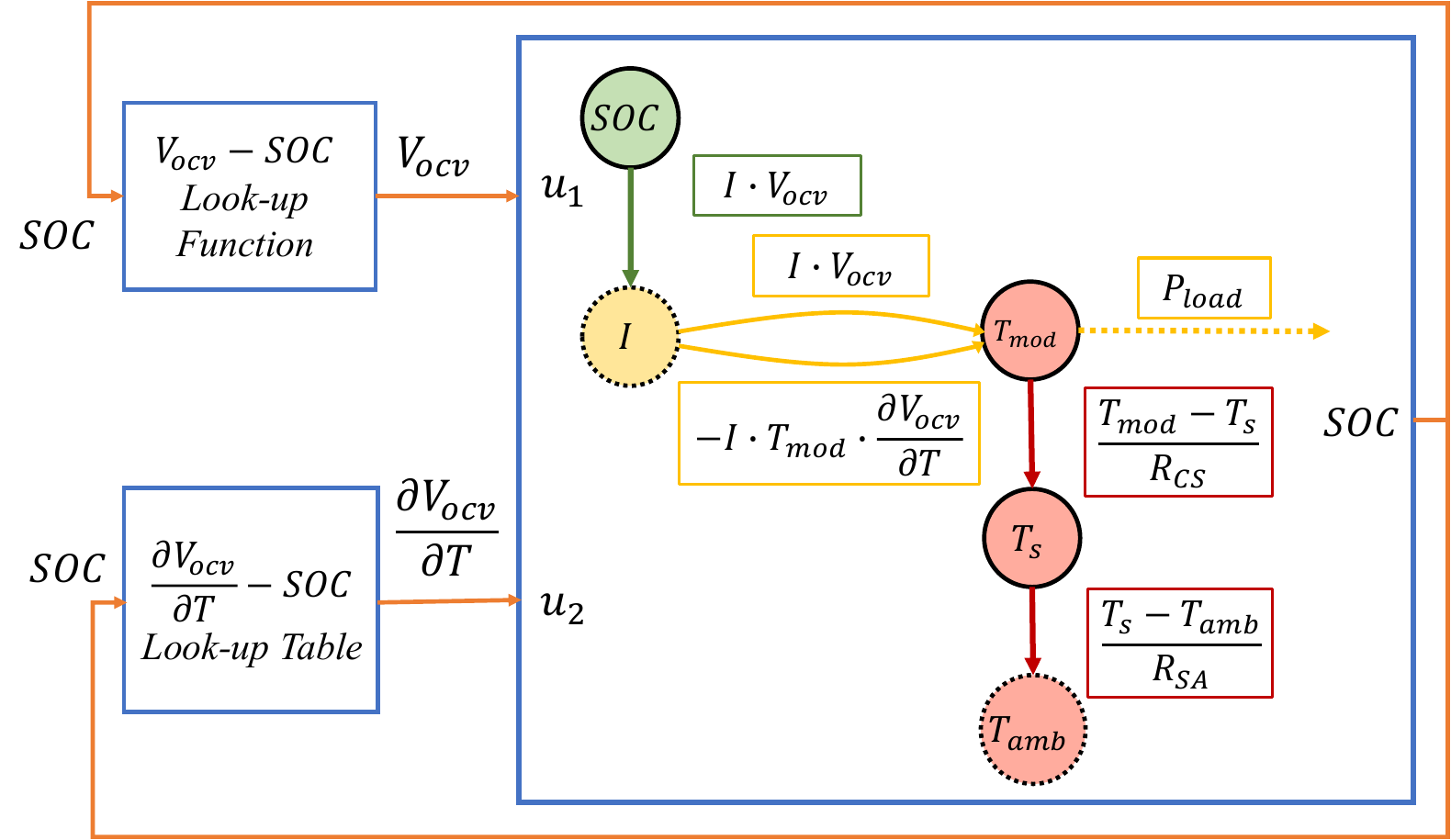}%
    \caption{Graph-based representation of the battery module.}
    \label{fig:module_graph}
\end{figure}

Using \eqref{eq:soc_dynamics_power} together with the thermal balances, the module is represented by the three dynamic vertices shown in Fig.~\ref{fig:module_graph}, with states
\[
x_1 = \mathrm{SOC}, \qquad
x_2 = T_{\mathrm{mod}}, \qquad
x_3 = T_s,
\]
and corresponding capacitances
\[
C_1 = {3600\,B_{\mathrm{cap,mod}}}V_{\mathrm{ocv,mod}}(\mathrm{SOC}),\]
\[
C_2 = C_{\mathrm{th,mod}}, \qquad
C_3 = C_{\mathrm{th,s}}.
\]

The electro-thermal interaction connecting the current source to the module-temperature vertex carries the total internal heat generation,
\begin{equation}
P =
I_{\mathrm{mod}}\,V_{\mathrm{ocv,mod}}(\mathrm{SOC})
-
I_{\mathrm{mod}}\,T_{\mathrm{mod}}
\left(\frac{\partial V_{\mathrm{ocv}}}{\partial T}\right)_{\mathrm{mod}}(\mathrm{SOC}),
\label{eq:total_heat}
\end{equation}
which is the only edge in the module model that requires the two methods introduced earlier.

Method~1 (recursive state-to-input feedback) is used because both $V_{\mathrm{ocv}}$ and $\partial V_{\mathrm{ocv}}/\partial T$ in \eqref{eq:total_heat} depend on the non-adjacent state $\mathrm{SOC}$. The SOC state is exposed as an output and mapped to the edge inputs
\[
u_{1,1} = V_{\mathrm{ocv,mod}}(\mathrm{SOC}), \qquad
u_{1,2} = \left(\frac{\partial V_{\mathrm{ocv}}}{\partial T}\right)_{\mathrm{mod}}(\mathrm{SOC}),
\]
so that the edge retains the edge structure \eqref{eq:generic_edge}.

Method~2 (parallel edge decomposition) is required because \eqref{eq:total_heat} contains two structurally different power-flow types, $x_t u_{1,1}$ and $x_t x_h u_{1,2}$. These cannot be represented by a single edge with one input. The interaction is therefore decomposed into two parallel edges,
\[
P_1 = I_{\mathrm{mod}} u_{1,1}, \qquad
P_2 = -I_{\mathrm{mod}}T_{\mathrm{mod}} u_{1,2},
\]
with the original heat generation recovered by $P = P_1 + P_2$. Both edges share the same head and tail vertices, and each conforms to the single-input edge form \eqref{eq:generic_edge}.

All remaining edge functions, which already match the edge structure \eqref{eq:generic_edge}, appear in Fig.~\ref{fig:module_graph} along with the complete graph model. The electrical power $P_{\mathrm{load}} = V_{T,\mathrm{mod}}\,I_{\mathrm{mod}},$ appearing in the $T_{\mathrm{mod}}$ balance represents the power demanded by the external system and is modeled as an external edge, since both $V_{T,\mathrm{mod}}$ and $I_{\mathrm{mod}}$ are measured. {Applying~\eqref{eq:vertex_dynamics} at each dynamic vertex recovers the original ODEs~\eqref{eq:Tmod_dynamics}--\eqref{eq:soc_dynamics_power}, verifying the graph model for the battery module.}

With the measured current $I_{\mathrm{mod}}$, ambient temperature $T_{\mathrm{amb}}$, and power load $P_{\mathrm{load}}$ provided as disturbances, the battery module dynamics can be simulated within the graph-based framework.

\subsection{Validation Results}

\begin{figure}[!t]
    \centering
    \includegraphics[width=\columnwidth]{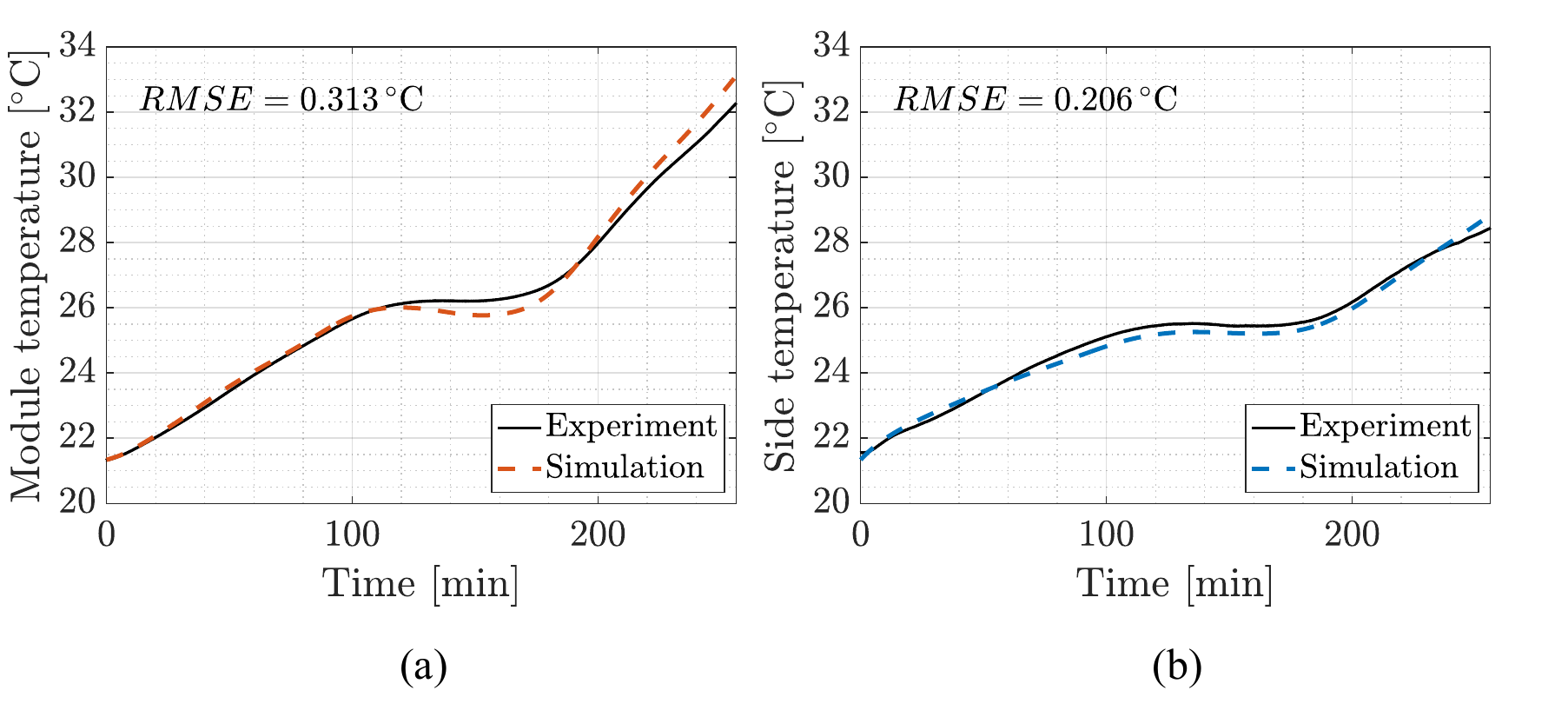}%
    \caption{Comparison of experimental and simulated module and side temperatures during a 30 A CC discharge.}
    \label{fig:batt_results}
\end{figure}

In the validation experiment, the full pack, including the chosen module, is discharged from near 100\% to about 20\% SOC under a constant-current (CC) load of 30~A at room temperature. The cooling fans are removed so that heat rejection is dominated by ambient passive cooling. The measured current, power load ($V_{T,\mathrm{mod}}\,.I_{\mathrm{mod}}$), and ambient temperature are used as disturbances to the graph-based battery module model, and the simulated temperatures are compared with the module temperature sensor measurements, as reported in Fig.~\ref{fig:batt_results}.
Fig.~\ref{fig:batt_results}(a) shows the internal module temperature $T_{\mathrm{mod}}$ and Fig.~\ref{fig:batt_results}(b) shows the side temperature $T_s$. In both cases, the graph-based model tracks the measurements closely over the discharge session, indicating that the graph-based formulation captures the module thermal behavior with sufficient fidelity while remaining compatible with the proposed modeling methods.

\section{Conclusion}
\label{sec:conclusion}

This paper extends graph-based modeling of multi-domain energy systems by introducing two general methods that address common limitations in prior formulations. A recursive state-to-input feedback scheme is formulated to represent power flows that depend on non-adjacent states while retaining the previously established edge structure and incidence representation. A parallel edge decomposition method is proposed for composite interactions in which a single physical connection carries multiple power flows driven by different inputs. The proposed methods are applied to construct an electro-thermal graph-based model of a Li-ion battery module, providing a concrete application of the extended framework.

Both methods preserve conservation of energy at the vertices and act locally at the component level, so components that use the recursive feedback or parallel edge constructions can be interconnected directly with components modeled using prior graph-based formulations without altering the system-level representation. As a result, the extended framework maintains the modularity and scalability of the graph-based approach for larger energy systems.

Future work will focus on applying the extended framework to larger and more complex energy systems, such as battery packs with active cooling, and on integrating the proposed methods with multi-state graph-based formulations to further expand the applicability of graph-based models.

\section*{DECLARATION OF GENERATIVE AI AND AI-ASSISTED TECHNOLOGIES IN THE WRITING PROCESS}
{During the preparation of this work, the author(s) used ChatGPT (OpenAI) to improve the English grammar and writing style of the manuscript. After using this tool, the author(s) reviewed and edited the content as needed and take(s) full responsibility for the content of the publication.}

\bibliography{ifacconf}
\end{document}